\documentstyle[aps]{revtex}
\begin{document}

\title{Quantum trajectories of interacting pseudo-spin-networks}
\author{C.M.~Granzow \and G.~Mahler}
\address{Institut f\"ur Theoretische Physik, Universit\"at Stuttgart, \\
Pfaffenwaldring 57, 70550 Stuttgart, Germany\\
(Fax: +49-711/685-4909, E-mail: claus@theo.physik.uni-stuttgart.de)}
\date{\today}
\maketitle
\begin{abstract}
We consider quantum trajectories of composite systems as
generated by the stochastic unraveling of the respective
Lindblad-master-equation. Their classical limit is taken to correspond
to local jumps between orthogonal  states. Based on
statistical distributions of jump- and inter-jump-distances we are
able to quantify
the non-classicality of quantum trajectories. To account for the
operational effect of entanglement we introduce the novel concept of 
``co-jumps".
\end{abstract}
\section{Introduction}
\label{intro}
In contrast to the ensemble description based on master equations,
quantum trajectories make available further information about details
which are lost by ensemble averaging. 
The study of quantum trajectories of open systems
\cite{carmichael}\cite{dalibard} 
has therefore found broad 
application in various fields to analyze fundamental processes.

It is important to distinguish the quantum trajectories in 
Hilbert space from the 
Feynman paths \cite{grabertschramm}: whereas the
latter are defined in the underlying classical phase space and
contribute with a complex probability amplitude to the path
integral, the Hilbert space quantum trajectories contribute  with a
real and positive probability to the ensemble density matrix
\cite{breuerpetruccione}.\\  
Hence in the Feynman path integral formulation the classical limit can
be obtained in a direct way when the interferences between the paths
basically reduce to the classical path.
In contrast the Hilbert space does not have a direct classical
analogue. Here, we define a Hilbert space-trajectory as ``classical",
if it is constrained to orthogonal states.
The jumps in the Hilbert space imply jumps in observable space like energy
or angular momentum: The orthogonal states can then be
interpreted as the eigenstates of some observable; its spectrum will
here taken to be discrete and finite (``telegraph signal"). 

While the non-classicality of states has attracted much interest
recently \cite{vedral,popescu,peres,werner,vogel}, the
non-classicality of the dynamical evolution, in 
particular on the level of trajectories, has received little if any
attention so far. It is tempting to expect that an increasing dissipative
interaction with the classical environment should make the
trajectories more and more ``classical". One may wonder, however,  to
what extent 
this expectation can be verified quantitatively. One may also wonder
whether there is a relationship between non-classicality of states and
of trajectories.

In this paper we introduce statistical distribution functions to
account for the non-classicality of quantum trajectories,
show their properties and their relation to the non-classicality of
states.  

This paper is organized as follows: In section
\ref{cluster-operators} we define a convenient operator set for the
description of quantum networks, the states of which will be discussed
in section \ref{quantum_states}. The unraveling of the Linblad master
equation into single quantum trajectories is summarized in section
\ref{stochastic_unraveling}. In section
\ref{measures_of_state_distance} we discuss some properties of the
measure of state distance, which are applied to the quantum
trajectories in section
\ref{state-distance_distributions_and_entanglement} and
\ref{co-jumps}. The concept of jump- and co-jump-distances is exemplified
in section \ref{co-jump_properties_of_model_states} for special two-, three-
and four-particle states. In section
\ref{jump_statistics_of_simulated_trajectories} we present numerical
results exploiting this jump-concept for simulated trajectories. We
conclude with a brief summary.

\section{Cluster-operators}
\label{cluster-operators}
We consider a network consisting of $N$ subsystems of $n$ states
each. The local states are $|p(\mu)\rangle$, $p=1,2,\dots,n$,
$\mu=1,2,\dots,N$, allowing to introduce the transition-operators
$\hat P_{pq}(\mu) = |p(\mu)\rangle\langle q(\mu)|$. These can be
combined to give the $n^2$ generators of the SU($n$)-algebra, which
read for $n=2$:
\begin{eqnarray}
\hat \lambda_1(\mu) &=& \hat P_{12} (\mu) + \hat P_{21} (\mu)\\
\hat \lambda_2(\mu) &=& i \left( \hat P_{12} (\mu) - \hat P_{21} (\mu)
\right) \\
\hat \lambda_3(\mu) &=& \hat P_{22} (\mu) - \hat P_{11} (\mu)\\
\hat \lambda_0(\mu) &=& \hat P_{11} (\mu) + \hat P_{22} (\mu) = \hat 1
(\mu) 
\end{eqnarray}
They constitute a complete, orthogonal set of $n^2-1$ traceless
operators. A corresponding set for the total network is then given by
the $n^{2N}$ product-operators: Here we will restrict ourselves to
$N=4$, $n=2$, in which case we have \cite{mahlerweberruss}
\begin{equation}
\hat Q_{mlkj}= \hat \lambda_m(4) \otimes \hat \lambda_l(3) \otimes
\hat \lambda_k(2) \otimes \hat \lambda_j(1) 
\end{equation}
The number $c$ of indices unequal zero is the number of subsystems
this operator acts on. There are $n_c={N \choose c} (n^2 -1 )^c$ such
$c$-cluster-operators, with $0\leq c \leq N$. Operators $\hat Q$
acting on different subsystems commute.
\section{Quantum states}
\label{quantum_states}
\subsection{Correlations}
Any network-operator in a given Liouville space $\{ N,n\}$ can be
expressed in terms of such cluster operators:
In particular, for the density-operator $\hat \rho$ one
finds (again for 
$N=4$, $n=2$, cf.\ Ref.\  \cite{mahlerweberruss})
\begin{equation}
\hat \rho = \frac{1}{2^4} \sum_{jklm} K_{mlkj} \hat Q_{mlkj}
\end{equation}
with $K_{mlkj} = \mbox{tr} \left\{ \hat \rho \hat Q_{mlkj}
\right\}$.\\
These expectation-values ${\bf K} = \left\{ K_{mlkj} \right\}$
uniquely specify the state; they decompose into $n_c$ $c$-point
correlation functions. The only $c$=0-term is $K_{0000} = \mbox{tr} \{
\hat \rho \} = 1$. The $c$=1-terms are the (local) Bloch-vectors,
$K_{000j} \equiv \lambda_j^{(1)}$, $K_{00k0} \equiv \lambda_k^{(2)}$
etc., which can be found from local (ensemble-)measurements.
The $c>1$-terms are typically inferred from
coincidence-measurements 
(ensemble measurements).
\subsection{Covariances}
\label{covariances}
One easily convinces oneself that these correlation functions $K_{mlkj}$
factorize if and only if the state $\hat \rho$ exhibits some product form:
For example, if $\hat \rho (4,3,2,1) = \hat \rho(4,2) \otimes \hat
\rho(3,1)$, then 
\begin{equation}
\label{cov1}
K_{mlkj} = K_{m0k0} \cdot K_{0l0j}
\end{equation}
or if $\hat \rho(4,3,2,1) = \hat \rho(4) \otimes \hat \rho(3,2)
\otimes \hat \rho(1)$, then 
\begin{equation}
\label{cov2}
K_{mlkj} = K_{m000} \cdot K_{0lk0} \cdot K_{000j} \quad \mbox{etc.}
\end{equation}
Subsystems and groups of subsystems which do {\em not} factor are
called ``entangled" ($\hat \rho$ of the total system is taken to be
pure and to describe a single network). Without any 
entanglement, all $c$-point correlation functions are thus reducible
to local expectation values (i.e.\ of type $c=1$).\\
It is therefore convenient to introduce state-parameters, which
describe deviations from this factorization property. For this purpose
we introduce a supplementary set of cluster-operators
\begin{equation}
\Delta \hat Q_{mlkj}= \Delta \hat \lambda_m(4) \otimes \Delta \hat
\lambda_l(3) \otimes \Delta \hat \lambda_k(2) \otimes \Delta \hat
\lambda_j(1)  
\end{equation}
based on the local ``deviation-operators" 
\begin{equation}
\Delta \hat \lambda_m(\mu) = \left\{ \begin{array}{ll}
\hat \lambda_m(\mu) - \lambda_m^{(\mu)} \hat 1 (\mu) & \mbox{for }
m \neq 0 \\ 
\hat 1 (\mu) & \mbox{for } m = 0 \end{array} \right.
\end{equation}
The respective expectation values (``quantum-covariances")
\begin{equation}
M_{mlkj} = \mbox{tr} \left\{ \Delta \hat Q_{mlkj} \hat \rho \right\}
\end{equation}
then also come in different $c$-classes: For $c=0$, $M_{0000}=K_{0000}=1$,
for $c=1$, $M_{000j}=0$ etc., for $c=2$, 
\begin{equation}
M_{00kj} = K_{00kj} - K_{00k0} \cdot K_{000j} \quad \mbox{etc.}
\end{equation}
for $c=3$,
\begin{eqnarray}
\nonumber
M_{0lkj} &=& K_{0lkj} - K_{0lk0} \cdot \lambda_j^{(1)} - K_{0l0j} \cdot
\lambda_k^{(2)}  \\
\label{m_0lkj}
&& - K_{00kj} \cdot \lambda_l^{(3)} +
2\lambda_l^{(3)}\cdot\lambda_k^{(2)}\cdot \lambda_j^{(1)} \quad
\mbox{etc.} 
\end{eqnarray}
The set of expectation values $\{ \lambda_k^{(\mu)} , M_{mlkj} \}$ can
alternatively be used to specify the network-state.
With all $\lambda_k^{(\mu)} =0$ we obviously have $M_{mlkj} = K_{mlkj}$.
The factoring properties of $K_{mlkj}$ carry over to $M_{mlkj}$. In
particular, under the condition as for eq.\ (\ref{cov2}) we get
$M_{mlkj} = M_{m000} \cdot M_{0lk0} \cdot M_{000j} = 0 $. In
general, any specific $M_{mlkj}$ is zero, if at least one individual
subsystem entering with a local operator-index $\neq 0$ factors out. 
\subsection{Entanglement measures}
\label{coherence_measures}
For a product state as of eq.\ (\ref{cov1}) we have $M_{mlkj} = M_{m0k0}
\cdot M_{0l0j} \neq 0 $.
By substracting all possible partitions $(c=c_1+c_2+ \dots$, $c_i \ge
2)$ we introduce (here for $c=4=2+2)$:
\begin{eqnarray}
\nonumber
\tilde M_{mlkj} = M_{mlkj} &-& M_{ml00}M_{00kj} \\  \nonumber
&-& M_{m0k0}M_{0l0j}\\ 
&-& M_{m00j}M_{0lk0}
\end{eqnarray}
which is thus zero for {\em any} product state.
For $c<4$ we obviously get $\tilde M = M$; for $c>4$ 
this connection scheme is easily generalized.\\
To quantify entanglement on the total network-level $(N=4)$ we use
(cf.\ Ref.\ \cite{mahlerweberruss})
\begin{equation}
\beta(4,3,2,1) = \sum_{m,l,k,j =1}^3 \left( \tilde M_{mlkj} \right) ^2
\end{equation}
as well as corresponding sub-space measures like 
\begin{equation}
\label{beta21}
\beta (2,1) = \sum_{k,j=1}^3 \left( \tilde M_{00kj} \right) ^2
\end{equation}
Note that $\beta(2,1) \neq 0$ indicates any entanglement $\tilde M_{00kj}
\neq 0$ between 
subsystems (2) and (1) only: $\beta(4,3,2,1)$ could still be zero.
These $\beta$-functions can be used instead of the ``entropies of
entanglement" (=entropy of the respective reduced density operators)
\cite{vedral,popescu}; the $M$-terms are easier to calculate and,
furthermore, can be made basis of approximation schemes (see below). 
\section{Quantum trajectories}
\subsection{Stochastic unraveling}
\label{stochastic_unraveling}
The evolution of open quantum systems is usually approximated by
the Lindblad Master equation, which is Markovian. Here, the influence
of the environment is specified by so called environment operators
$\hat L_s$ and the corresponding damping rates $W_s$:
\begin{eqnarray}
\nonumber
\frac{{\sl \partial}}{{\sl \partial} t} \hat \rho  +
\frac{i}{\hbar} [\hat H , \hat \rho ]  &=& \sum_s(
- \frac{1}{2} W_s \{ 
\hat L^+_s \hat L_s \hat \rho + \hat \rho \hat L^+_s \hat L_s \}\\
&&+ 
W_s \hat L_s \hat \rho \hat L^+_s)
\end{eqnarray}
Those operators $\hat L_s$ play a crucial role in the stochastic
unraveling of the master equation. The coupling to the environment
leads to individual quantum 
jumps between which there is a non-unitary continuous evolution
\cite{carmichael}.  
The jumps are generated by the last term of the right-hand side whereas
the first two terms can be combined with the Hamilton operator into a
non-Hermitian effective Hamiltonian responsible for the continuous
inter-jump-evolution. The probability for a quantum jump of type $s$
after the time interval $\delta t$ is given by 
\begin{equation}
p_s = W_s \mbox{tr} \{ \hat L_s \hat \rho \hat L_s^+ \} \delta t
\end{equation}

We suppose that these projections can be expressed in
terms of, in general, non-Hermitian operators, $\hat L_s = \hat
P_s(\mu)$ which are taken to act locally on one of the subsystems
$(\mu)$, 
\begin{equation} 
\label{proj}
\hat \rho ' =  \frac{\hat P_s(\mu) \hat \rho \hat P_s(\mu)^+
}{\mbox{tr}\left\{ \hat P_s(\mu) \hat \rho \hat P_s(\mu)^+
\right\} }
\end{equation}
For the characterization of these pure-state-trajectories the timing
of jumps plays 
a central role because it may give rise to measurable events
and to count- and waiting-time-statistics \cite{wawermahler}.
Unfortunately, however, these give only indirect evidence for the
non-classicality of trajectories (like, e.g., anti-bunching). We therefore
propose to supplement the analysis by directly referring to the motion
in Liouville-space. 

The characterization of single quantum trajectories with respect to
classicality measures could be done in different ways. In addition to the
possibilities presented below one may think of calculating the 
distribution functions of local as well as non-local coherence
measures (cf.\ Sect.\ (\ref{coherence_measures})). However, it should
be noted that these properties 
depend on the basic operators chosen for the state description. In
the case that the coupling to the environment leads to the built-up of 
states which do not happen to coincide with eigenstates of the local
operators $\hat \lambda_3(\mu)$, this approach will not show a
classical limit (i.e.\ $\alpha$ and/or $\beta$ remain unequal zero). 
\subsection{Measures of state distance}
\label{measures_of_state_distance}
There have been different proposals for defining a metric for
density matrices (see, e.g.~the Bures metric \cite{huebner}, fidelity
\cite{josza} or mutual information \cite{vedral}). For the
non-orthogonal Glauber-states $|\alpha \rangle$, e.g., a ``distance"
$d$ has been proposed \cite{brune} with $|\langle \alpha | \alpha '
\rangle |^2 = \exp{\{-d^2\}}$. In this
paper we use a measure, $D$, for the distance between two arbitrary
(generally mixed) states $\hat \rho$ and $\hat \rho'$ according to 
\begin{equation} D^2_{\hat \rho \hat \rho '} = \mbox{tr} \{ (\hat \rho - \hat
\rho')^2\} 
\end{equation}
which is, independent of the
dimension of the Liouville space, between 0 and 2. 
The maximum (squared) distance of 2 applies to orthogonal states. In
the case of pure states $(\hat \rho = |\Psi \rangle \langle \Psi |)$
$D^2_{\hat \rho \hat \rho '}$ can be rewritten 
as :
\begin{equation} 
\label{d2psi}
D^2_{\Psi \Psi'} = 2(1-|\langle \Psi | \Psi ' \rangle | ^2 ) \; . 
\end{equation}
It is easy to show, that $D_{\hat \rho \hat \rho ' }$ satisfies the
metric properties \cite{reedsimon} in the Liouville space $\cal L$, i.e.\ 
\begin{eqnarray}
\nonumber
D_{\hat \rho \hat \rho ' } &\ge& 0 \quad \mbox{for all $\hat \rho$,
$\hat \rho'$ of $\cal L$,}\\ \nonumber
D_{\hat \rho \hat \rho ' } &=& 0  \quad \mbox{if and only if $\hat
\rho = \hat \rho '$,}\\ \nonumber
D_{\hat \rho \hat \rho ' } &=& D_{\hat \rho' \hat \rho } \quad \mbox{
for all $\hat \rho$, $\hat \rho'$ of $\cal L$,}\\
D_{\hat \rho \hat \rho ' } &\le& D_{\hat \rho \hat \rho '' } + D_{\hat
\rho'' \hat \rho ' } \quad \mbox{ (triangle inequality).}
\end{eqnarray}
This measure $D_{\hat \rho \hat \rho ' }^2$ can directly be
expressed in terms of the SU(2)-parameters, $K_{mlkj}$, namely as the
squared length of the difference vector between 
\boldmath ${\bf K}$\unboldmath$=\{K_{mlkj}\}$ and \boldmath${\bf
K'}$\unboldmath$=\{K_{mlkj}'\}$, 
\unboldmath
\begin{equation}
\label{distsun}
D_{\hat \rho \hat \rho ' }^2 
= \frac{1}{2^4} \sum_{j,k,l,m=0}^3 \left( K_{mlkj} - K_{mlkj}'
\right)^2 \\  \nonumber
\end{equation}
This concept of state distance can easily be generalized to distances
defined on reduced state spaces: Observing that, e.g.,\
\begin{equation}
\mbox{tr}_{\{4,3,2\}} \left\{ \hat Q_{mlkj} \right\} = \hat
\lambda_j(1) 2^3 \delta_{m0} 
\delta_{l0} \delta_{k0}
\end{equation}
(here $\mbox{tr}_{\{\mu\}}$ means trace-operation within $\mu$-subspace
only), we find for the reduced density operator of subsystem (1), say,
\begin{equation}
\hat \rho (1) = \mbox{tr}_{\{432\}} \hat \rho = \frac{1}{2} \sum_j
K_{000j} \hat \lambda_j (1)
\end{equation}
and
\begin{equation}
\label{reddis1}
\left( D^{(1)}_{\{ \hat \rho , \hat \rho' \}} \right)^2 = \frac{1}{2}
\sum_j \left( K_{000j} - K_{000j}' \right) ^2 
\end{equation}
Correspondingly, the distance as seen from the subsystems $(4,3,2)$ is
\begin{equation}
\label{reddis2}
\left( D^{(4,3,2)}_{\{ \hat \rho , \hat \rho' \}} \right)^2 = \frac{1}{2^3}
\sum_{k,l,m} \left( K_{mlk0} - K_{mlk0}' \right) ^2 
\end{equation}
Then we have by inspection the inequality 
\begin{equation}
\frac{1}{2^3} \left( D^{(1)}_{\{ \hat \rho , \hat \rho' \}} \right)^2 + 
\frac{1}{2} \left( D^{(4,3,2)}_{\{ \hat \rho , \hat \rho' \}}
\right)^2  \leq  D^2_{\{ \hat \rho , \hat \rho' \}} 
\end{equation}

\subsection{State-distance distributions}
\label{state-distance_distributions_and_entanglement}
In order to characterize quantum trajectories we introduce various
types of state-distances:  
the ``jump distance", by inserting into eq. (\ref{d2psi}) the
state right before and after the jump (in analogy to
the jump distance of the Brownian motion in classical physics
\cite{vankampen}), the ``inter-jump-distance" as the distance
between the final state of the last jump and the initial state of the
following jump, and the state distance
for a specified time interval $\tau$ during the
evolution of a given quantum trajectory,
\begin{equation} D^2_{\tau} = \mbox{tr} \{ \left( \hat \rho (t) -
\hat \rho (t+\tau) \right)^2 \} .
\end{equation}
Finally, we will be interested not only in the jump distance of the total
system, but also of parts of the system in their respective reduced
space $\mu$. Therefore we use 
$\left( D^{(\mu)}_{\{\hat \rho \hat \rho '\}} \right) ^2$ as defined in
eq.\ (\ref{reddis1},\ref{reddis2}) where $\hat \rho$ ($\hat \rho '$)
is the total density operator 
before (after) jump. In this way we can test to what extent a 
projection in subspace $(\nu; \nu \neq \mu)$ affects the reduced
state of subsystem $\mu$ (``co-jumps").

Sampling over one individual trajectory we find the
corresponding distribution functions, $f(D^2)$. These are normalized: 
\begin{equation}
\int_0^2 f(D^2) \, d(D)^2 =1 .
\end{equation}

\subsection{Co-jumps and entanglement}
\label{co-jumps}

One easily shows that for any complete (POVM)-type measurement
\cite{nielsen}, the {\em  ensemble-averaged} co-jump must be zero: 
For this purpose we write, for the (ensemble) density-operator after
measurement in (2,1), e.g.\ 
\begin{equation}
\hat \rho ' = \sum_s \hat P_s(2,1) \hat \rho \hat P_s^+(2,1)
\end{equation}
with $\sum_s \hat P_s^+(2,1) \hat P_s(2,1)=\hat 1$
and consider (overlines indicate ensemble-averaging)
\begin{eqnarray}
\overline{K}_{ml00}' &=& \mbox{tr} \left\{ \hat \rho ' \hat Q_{ml00} \right\}\\
&=& \mbox{tr} \left\{ \sum_s \left( \hat P_s (2,1) \hat \rho \hat
    P_s^+(2,1) \hat Q_{ml00} \right) \right\}\\
&=& \mbox{tr} \left\{ \sum_s \left( \hat P_s^+(2,1)\hat P_s (2,1)
  \right) \hat  \rho  \hat Q_{ml00}  \right\}\\ 
&=& \mbox{tr} \left\{ \hat  \rho  \hat Q_{ml00}  \right\} =
\overline{K}_{ml00}
\end{eqnarray}
Here we have made use of the fact that $\hat P_s^+(2,1)$ and $\hat
Q_{ml00}$ commute as they act on different sub-spaces. We conclude
that 
\begin{equation}
\label{dkquer}
 \left( D^{(4,3)}_{\{\hat \rho \hat \rho ' \} } \right) ^2 = \frac{1}{2^2}
\sum_{ml} \left( \overline{K}_{ml00} - \overline{K}_{ml00}' \right)^2
= 0 
\end{equation}
This means that ensemble quantum mechanics is local in an operational
sense: Measuring in some subspace, here (2,1), does not have any
influence outside this subspace. An analogue statement holds for
respective unitary transformations. 

However, as is known since the famous EPR-experiments \cite{aspect}
individual measurements leading to new 
information may violate this locality.
To show this we consider
\begin{equation}
\hat \rho ' = \frac{1}{p_s} \hat P_s(2,1) \hat \rho \hat P_s^+ (2,1)
\end{equation}
where $p_s=\mbox{tr}\left\{ \hat P_s^+(2,1) \hat P_s (2,1) \hat \rho
\right\}$. Based on the same arguments as before we obtain
\begin{eqnarray}
\nonumber
K_{ml00}' = K_{ml00} + \frac{1}{p_s} & & \mbox{tr} \Big\{ \hat \rho
  \left( \hat Q_{ml00} - \hat 1 K_{ml00} \right) \\ 
& & \times \hat P_s^+ (2,1) \hat P_s (2,1) \Big\}
\end{eqnarray}
Now let $\hat P_s^+ \hat P_s = \frac{1}{2} \left( \hat 1 (1) + \hat
  \lambda_3 (1) \right) = \hat P_{22} (1)$: Then we get
\begin{equation}
K_{ml00}'-K_{ml00} = \frac{1}{2p_s} \left( K_{ml03} - K_{ml00} \cdot
      K_{0003} \right) 
\end{equation}
The right hand side is zero if $K_{ml03} = K_{ml00} \cdot K_{0003}$,
i.e.\ if subsystem (1), which is measured, has no entanglement with
the subsystem (4,3). Otherwise,
\begin{equation}
\left(D^{(4,3)}_{\{\hat \rho \hat \rho ' \} }\right)^2 = \frac{1}{2^2}
\sum_{ml}  
\left( K_{ml00} - K_{ml00}' \right)^2  \neq 0
\end{equation}
which is thus non-zero also for the ensemble: Comparing with eq.\
(\ref{dkquer}) we note that
\begin{equation}
\overline{\left( K_{ml00} - K_{ml00}' \right)^2} \neq
\left( \bar K_{ml00} - \bar K_{ml00}' \right)^2 .
\end{equation}

\section{Co-jump properties of model states}
\label{co-jump_properties_of_model_states}
Our intention is to use the concept of co-jumps for the
characterization of single quantum trajectories. As these trajectories
always connect pure states, ``mixed states" only appear 
for reduced subspaces. Any non-zero
entropy of such reduced density matrices is due to 
entanglement and not due to our incomplete knowledge.
Co-jumps will occur also within reduced spaces;
such situations will be included below in a formal way. 

In the following we perform
individual ``measurements" based on the specific operators (cf.\ eq.\
(\ref{proj})) 
\begin{equation}
\hat P_{i}(\mu) = \frac{1}{2} \left( \hat
1(\mu) + \hat 
\lambda_{i}(\mu) \right)   , i=0,1,2,3 \; . 
\end{equation}
Note that these measurements correspond to only one outcome each.
In the case of coincidence measurements we take the dyadic product
of such single particle operators. 
According to eq.\ (\ref{distsun}) and eq.\
(\ref{reddis1},\ref{reddis2}) we then calculate the 
jump- and co-jump-distance.

\subsection{Two-particle state}
\label{twopart}
\begin{figure}[h]
   \setlength{\unitlength}{1cm}
   \begin{picture}(15,5)
      \put(0,-13){
         \includegraphics{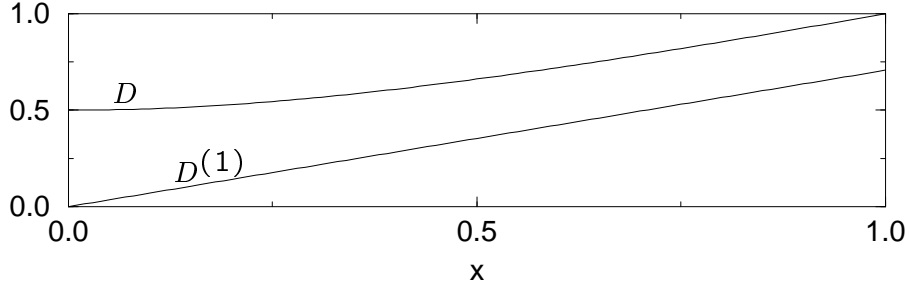}
      }
      \put(8.5,2.5){}
   \end{picture}
\caption[]{Co-jump $D^{(1)}$ and total jump $D$ for the
Werner-state $\hat \rho_x$ which is projected by $\hat
P_i^{(2)}$ 
($x=1 \to$ EPR-state)}
\label{fig1}
\end{figure}

We first consider the completely mixed state, the EPR-state
(``cat-state") $ | \mbox{EPR} \rangle=\frac{1}{\sqrt{2}} \left( | 12
\rangle - | 21 \rangle 
\right) $, and the ``Werner-state" \cite{werner}
\begin{equation}
\label{werner}
\hat \rho_x = (1-x) \frac{1}{4} \hat 1 + x 
| \mbox{EPR} \rangle \langle \mbox{EPR} |  \quad ,  0 \leq x \leq 1 \; .
\end{equation}
The maximum entanglement for the latter is reached for $x=1$
(EPR-state, $\beta (2,1)=3$).  
Table \ref{tab1} shows results for $x=0$ and $x=1$, conditioned by the
respective projection, $i$. Jump and co-jump are independent of $i$,
confirming the ``isotropy" of the EPR-state and of the mixed state.\\
In Fig.\ \ref{fig1} one can see the transition from the 
mixed state to the EPR-state. In contrast to $D$, the
co-jump $D^{(1)}$ is linear in the whole region from
$x=0$ to $x=1$: The state (\ref{werner}) reacts in a local way for
$x=0$ only. 

The non-locality, to be sure, could still be ``explained" by a
local hidden-variable theory \cite{popescu} unless $x>1/\sqrt{2}$
(violation of Bell-inequalities) or $x>1/3$ (violation of separability
condition \cite{peres}), respectively.
These interpretations have to assume, though, that mixed states can be
treated as classical mixtures (i.e.\ resulting from our ignorance
\cite{cohen} rather than from undefined properties due to
entanglement with other subsystems). Stochastic modeling deals with
single networks and individual subsystems (additional knowledge!);
co-jumps within the latter then reflect their (``objective") change as
induced by a distant observation.

\begin{table}[h]
\label{tab1}
\caption[ ]{(Squared) jump distances for the specific $N$=2-particle
state $\hat \rho_x$ (see eq.\ (\ref{werner}))
(projection by $\hat P_i^{(2)}$)} 
\begin{tabular}{llll}
\hline\noalign{\smallskip}
x & $i$ & $ \left( D^{(1)} \right) ^2$ & $(D)^2$ \\
\noalign{\smallskip}\hline\noalign{\smallskip}
1 & 1 & 0.5 & 1.0 \\
1 & 2 & 0.5 & 1.0 \\
1 & 3 & 0.5 & 1.0 \\
0 & 1 & 0.0 & 0.25 \\
0 & 2 & 0.0 & 0.25 \\
0 & 3 & 0.0 & 0.25 \\
\noalign{\smallskip}\hline
\end{tabular}
\end{table}

\subsection{Three-particle state}
As a representative for the $N$=3-case we consider the cat-state
(GHZ-state, $\beta(3,2,1)=4$, $\beta(2,1)=1$)
\begin{equation}
|\mbox{GHZ} \rangle = \frac{1}{\sqrt{2}} \left( |111\rangle + |222 \rangle
\right) 
\end{equation}
 and study projections on particle 3, and on particles 2 and
3 in coincidence. In contrast to the EPR-state, the GHZ-state is not
rotationally invariant, the $i$=3-direction plays a special role which can
be seen by comparing the result for $(i,j)=(0,1)$ and $(0,3)$ or
$(1,1)$, $(3,3)$ (see table \ref{tab2}). 
\begin{table}
\label{tab2}
\caption[ ]{(Squared) jump distances for the $N$=3-cat-state (projection by $\hat
  P_i^{(2)} \otimes \hat P_j^{(3)}$)} 
\begin{tabular}{lllll}
\hline\noalign{\smallskip}
 $i$ & $j$ & $ \left( D^{(1)} \right)^2$ & $ \left( D^{(1,2)}
 \right)^2$ & $(D)^2$ \\ 
\noalign{\smallskip}\hline\noalign{\smallskip}
 0 & 1 & 0.0 & 0.5 & 1.0 \\
 0 & 2 & 0.0 & 0.5 & 1.0 \\
 0 & 3 & 0.5 & 0.5 & 1.0 \\
 1 & 1 & 0.5 & - & 1.5 \\
 1 & 2 & 0.5 & - & 1.5 \\
 1 & 3 & 0.5 & - & 1.5 \\
 3 & 1 & 0.5 & - & 1.5 \\
 3 & 2 & 0.5 & - & 1.5 \\
 3 & 3 & 0.5 & - & 1.0 \\
\noalign{\smallskip}\hline
\end{tabular}
\end{table}
One should note that the reduced state $\hat \rho (3,1)$ could be
written as a 
mixed product state (i.e.\ separable in the sense of Ref.\
\cite{peres}), which, nevertheless, leads to a co-jump in (1) induced
by subsystem (3). Only without further knowledge could this effect
be interpreted as being due to classical correlations,  cf.\ Sect.\
\ref{twopart}. 
\subsection{Four-particle state}
The special four-particle ``cat-state" ($\beta(4,3,2,1)=12$,
$\beta(3,2,1)=0$, $\beta(2,1)=1$)
\begin{equation}
|\Psi_{Cat} \rangle = \frac{1}{\sqrt{2}} \left( |1111 \rangle + |2222 \rangle
\right)
\end{equation}
shows a co-jump behavior which is, contrary to the $N$=3-cat state,
different also for $(i,j)=(1,1)$ and 
$(1,3)$. The pertinent co-jump properties of this
state are summarized in table \ref{tab3}.

\begin{table}
\label{tab3}
\caption[ ]{(Squared) jump distances for the $N$=4-cat-state
(projection by $\hat P_i^{(3)}\otimes 
\hat P_j^{(4)}$)} 
\begin{tabular}{lllll}
\hline\noalign{\smallskip}
 $i$ & $j$ & $ \left( D^{(1)} \right)^2$ & $ \left( D^{(1,2)}
 \right)^2$ & $(D)^2$ \\ 
\noalign{\smallskip}\hline\noalign{\smallskip}
 0 & 1 & 0.0 & 0.0 & 1.0 \\
 0 & 2 & 0.0 & 0.0 & 1.0 \\
 0 & 3 & 0.5 & 0.5 & 1.0 \\
 1 & 1 & 0.0 & 0.5 & 1.5 \\
 1 & 2 & 0.0 & 0.5 & 1.5 \\
 1 & 3 & 0.5 & 0.5 & 1.5 \\
 3 & 1 & 0.5 & 0.5 & 1.5 \\
 3 & 2 & 0.5 & 0.5 & 1.5 \\
 3 & 3 & 0.5 & 0.5 & 1.0 \\
\noalign{\smallskip}\hline
\end{tabular}
\end{table}
\section{Jump statistics of simulated trajectories}
\label{jump_statistics_of_simulated_trajectories}
\subsection{Hamilton-model}
In our simulations we consider an open  quantum network consisting of pseudo
spins (two-level systems) which interact with each other, with
external electro-magnetic fields and the environment. In rotating wave
approximation the 
local Hamiltonian of node $\mu$ can be expressed as ($\mu=1,2,3,4$) 
\begin{equation}
 \hat H(\mu) = \frac{1}{2} \; \delta^{\mu} \; \hat \lambda_3(\mu)+
\frac{1}{2} \; g^{\mu} \; \hat \lambda_1(\mu) \; .
\end{equation}
The external fields are characterized by the Rabi-frequencies
$g^{\mu}$ and the energy difference between the laser photons and the
spin energy is denoted by $\delta^{\mu}$. \\
The interaction between the spins will be given by the non-resonant
coupling, 
\begin{equation}
\hat H(\mu,\nu) = -\frac{1}{2} C_R^{(\mu,\nu)}  \; \hat \lambda_3(\mu)
\otimes \hat 
\lambda_3(\nu) \; ,
\end{equation}
which leads to an energy shift of one subsystem depending on
the state of the other.
In the following we will call a network ``homogeneous", if all model
parameters (including those for the bath coupling) are invariant under
any subsystem-index permutation. 

\subsection{Bath model}
In addition to the Hamiltonian model, we have to specify the coupling
to the environment leading to decoherence effects. For simplicity we
will take into account only dissipation of energy into the bath with
rate $W_{\downarrow}^{\mu}$ and operator $ \hat L_{\downarrow}(\mu) =
\hat P_{12}(\mu) $ and for the case of a bath with finite
temperature also excitations out of the bath with rate
$W_{\uparrow}^{\mu}$ and operator $ \hat L_{\uparrow}(\mu) =  \hat
P_{21}(\mu)$ 
 
In the case of a bath temperature $T=0$ there are only jumps into
the ground state, from where only the coherent laser field can drive
the state out again. Therefore the distribution functions for jump-
and inter-jump distances are virtually identical. 
\begin{figure*}
   \setlength{\unitlength}{1cm}
   \begin{picture}(15,6)
      \put(-2.5,7){
         \includegraphics{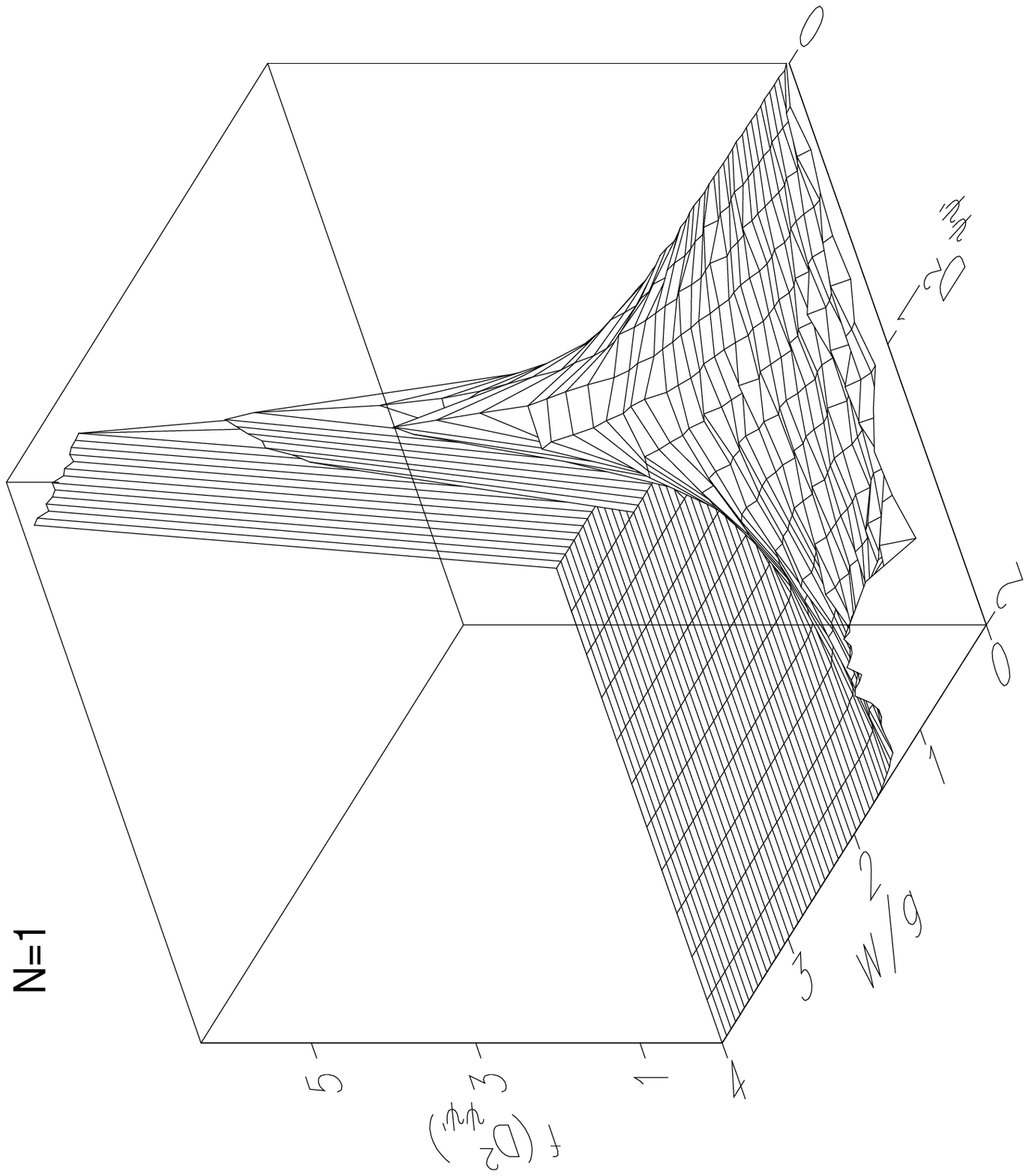}
      }
      \put(3.5,7){
         \includegraphics{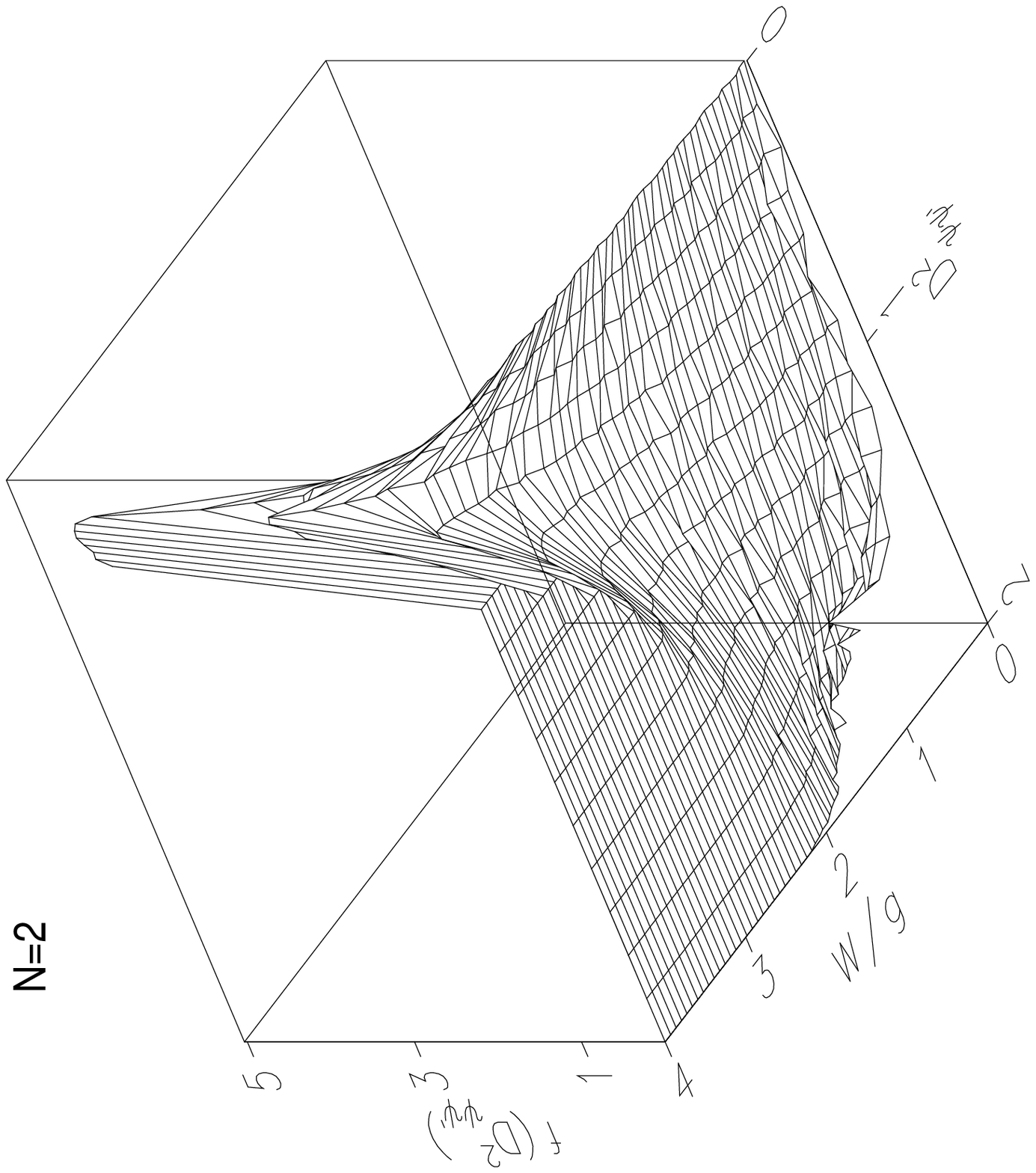}
      }
      \put(9.5,7){
         \includegraphics{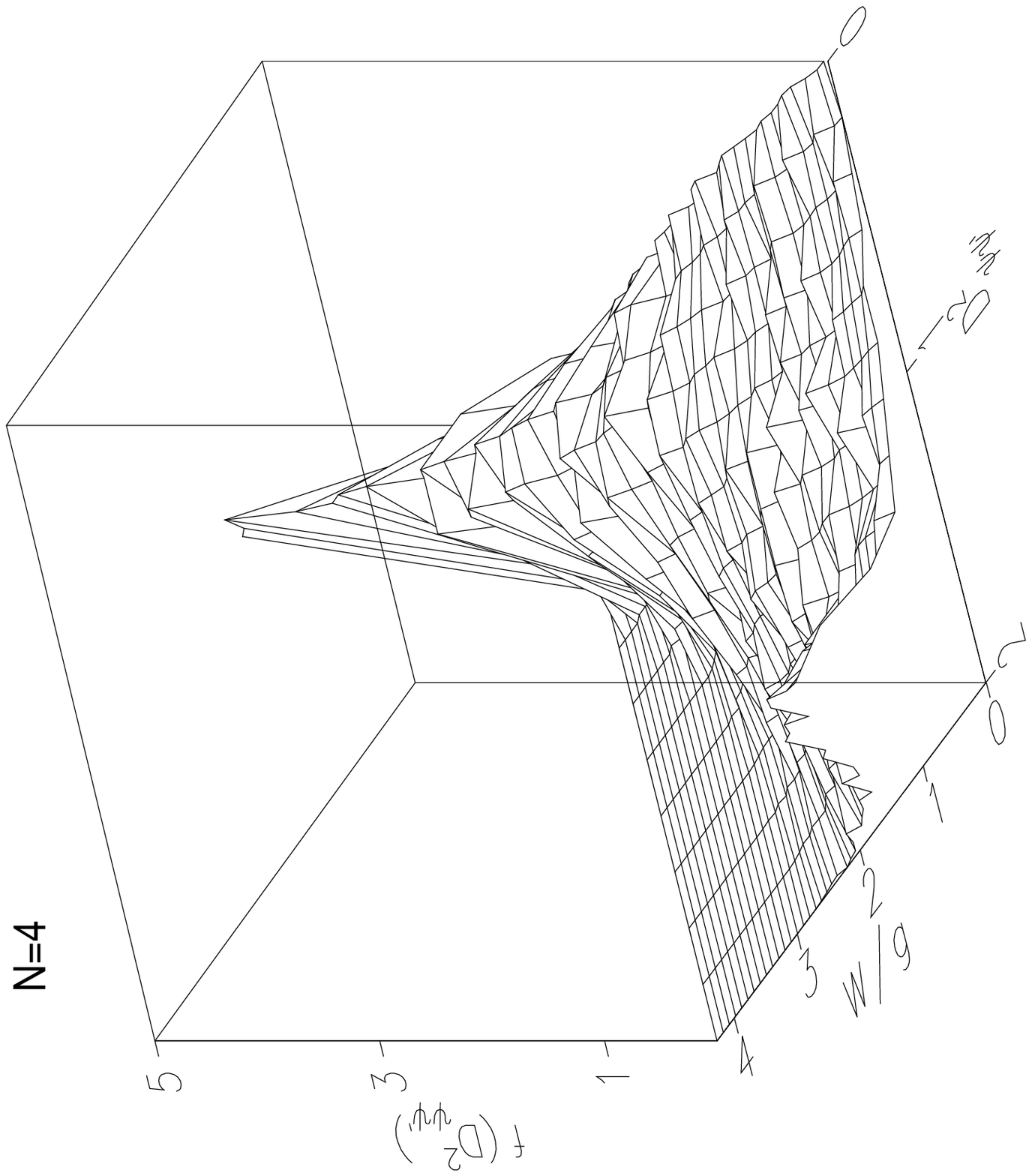}
      }
      \put(8.5,2.5){}
   \end{picture}
\caption[]{Distribution function $f(D^2_{\Psi \Psi'})$ for the jump
distance $D$ of a homogeneous network $N$ in interaction with a
bath at $T=0$ (damping rate $W_{\downarrow}^{\mu}=W$, $g^{\mu}=g=$ coupling
parameter to the coherent driving field, coupling $C_R^{(\mu,\nu)}
= 20$, detuning $\delta^{\mu} = -20$ for $N=2$ $(-40$ for $N=4$) )} 
\label{fig2}
\end{figure*}
For small environmental influence the broad distribution reaches its maximum
at $D^2_{\Psi \Psi'} = 2$, due to the fact, that 
jumps (photon-emissions) most likely start from the upper level in
which case the jump into the ground level has (squared)
distance 2. Increasing the ratio $W/g$ raises the
probability for jumps already for incomplete excitation,
which results in a (squared) jump distance smaller than 2. In the high 
damping limit, the jump distance peaks at zero and the
distribution becomes narrow (see Fig.\ \ref{fig2}).

The jump-distribution function $f(D^2)$ for homogeneous networks
shows a 
scaling behavior which is only weakly dependent on size
$N$ (Fig.\ \ref{fig2}). Contrary to naive
expectation, the limit $N \to \infty$ thus does not necessarily mean a
classical limit. For $W/g \gg 1$ the dynamics is virtually suppressed.

\begin{figure}
   \setlength{\unitlength}{1cm}
   \begin{picture}(15,6)
      \put(-2,7){
         \includegraphics{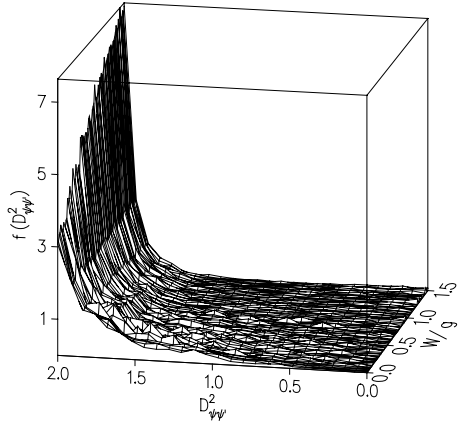}
      }
      \put(8.5,2.5){}
   \end{picture}
\caption[]{Distribution function $f(D^2_{\Psi \Psi'})$ for the 
jump distance $D$ for a network $N=2$ in interaction with a
bath of high temperature (damping rate $W_{\uparrow}^{\mu} =
W_{\downarrow}^{\mu} =W $, 
$g^{\mu}=g=$ coupling parameter to the coherent driving field,
renormalisation coupling $C_R= 20$, detuning $\delta^{\mu} = -20$)  }
\label{fig3}
\end{figure}

\begin{figure}
   \setlength{\unitlength}{1cm}
   \begin{picture}(15,6)
      \put(-2,7){
         \includegraphics{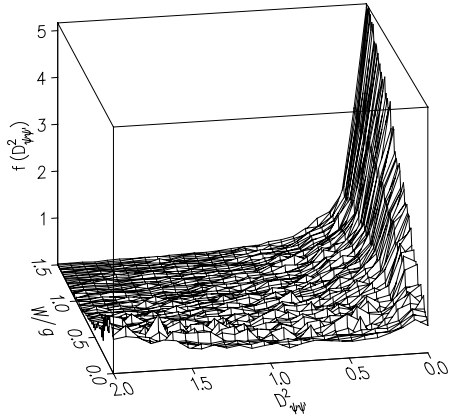}
      }
      \put(8.5,2.5){}
   \end{picture}
\caption[]{Distribution function $f(D^2_{\Psi \Psi'})$ for the inter
jump  distance, $D$, for a network $N=2$ in interaction with a
bath of high temperature (parameters see Fig.\ \ref{fig3})}
\label{fig4}
\end{figure}

The distribution changes completely, if the bath also
causes excitations (bath temperature $T \neq 0$):
Now, classical trajectories result, if the jump distance distribution
peaks at $D^2 = 2$ and becomes 
very narrow, while the inter-jump-distances tend to zero
(Fig.\ \ref{fig3} and Fig.\ \ref{fig4}). This happens
for $W/g \gg 1$; it will always happen for incoherent driving.


\subsection{Influence of entanglement}
We may
truncate the expansion of correlation functions at a certain
level, $c$, i.e.\ we neglect all orders of entanglement $\tilde M_{mlkj}$
above this 
specified level. This leads to a reduction of relevant state
parameters (cf.\ Sect.\ \ref{covariances}), but at the same time to
non-linear evolution equations. (For $c=2$, we replace, e.g.\
$K_{0lkj}$ by the various factors as obtained from eq.\ (\ref{m_0lkj}) with
$M_{0lkj}=0$.)
As this factorization does not correspond to a concrete
partitioning, those equations are not guaranteed to remain consistent: 
To the contrary, pure states (of closed systems) do no longer remain
pure, so that quantum 
trajectories based on this strategy lack a definite interpretation
as a single system quantum evolution. Only in the presence of sufficient
damping such deficiencies are negligible.

Alternatively, we may
carry out simulations for which the state is taken to factor into
concrete partitions, like, e.g., $\hat \rho (4,3,2,1) = \hat \rho (4,3)
\otimes \hat \rho (2,1)$. As a consequence, all entanglement terms,
$\tilde M_{mlkj}$, between those partitioned subgroups disappear. 
In this case the equations of motions remain consistent, i.e.\ the
density matrix for a closed system keeps its desired properties and a
pure state stays 
pure. The comparison of these simulations with the exact ones  
then allows to isolate the effect of various entanglements $\tilde M_{mlkj}$
on the level of individual quantum trajectories. 

\begin{figure}
   \setlength{\unitlength}{1cm}
   \begin{picture}(15,6)
      \put(-2,7){
         \includegraphics{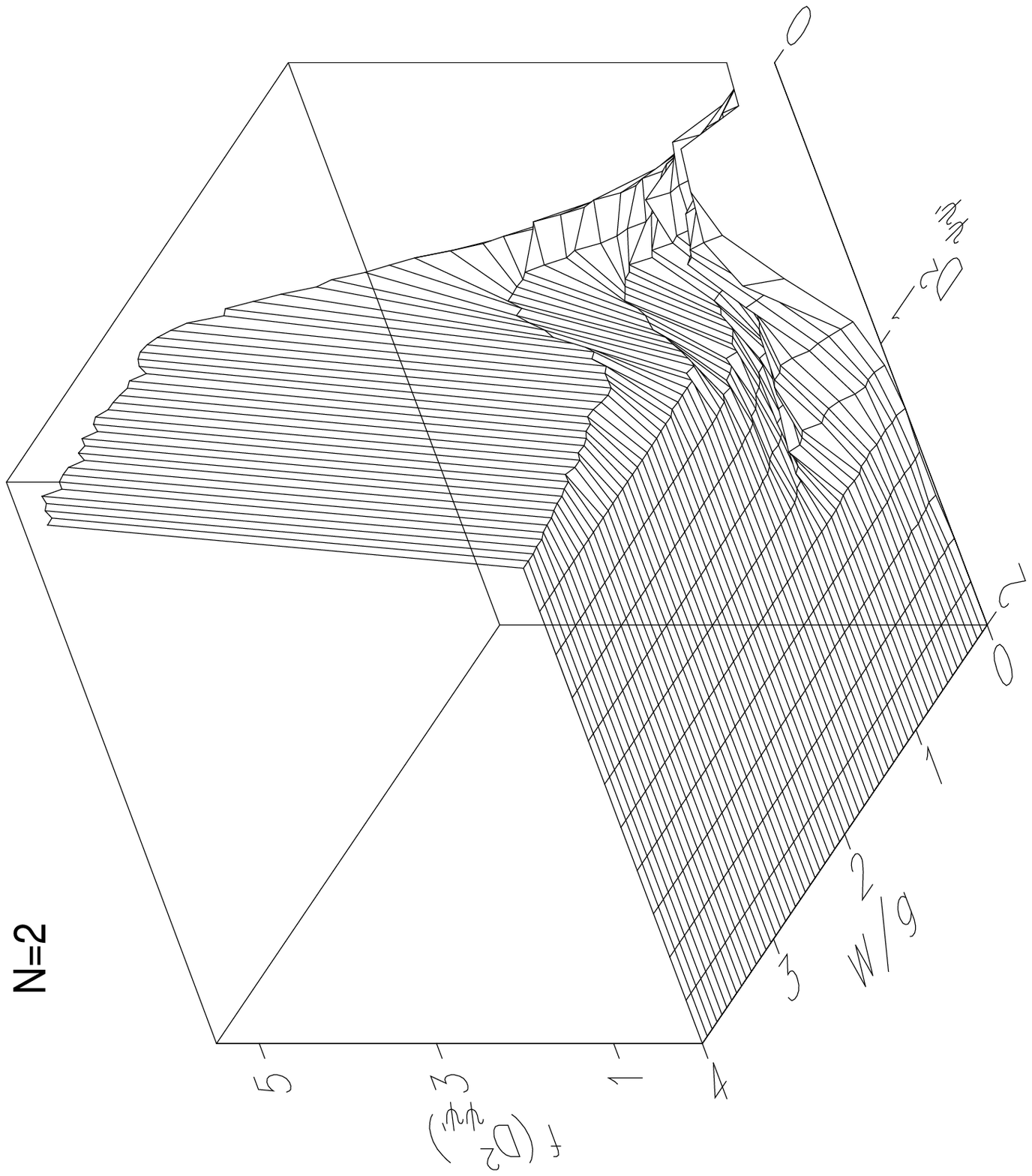}
      }
      \put(6,7){
         \includegraphics{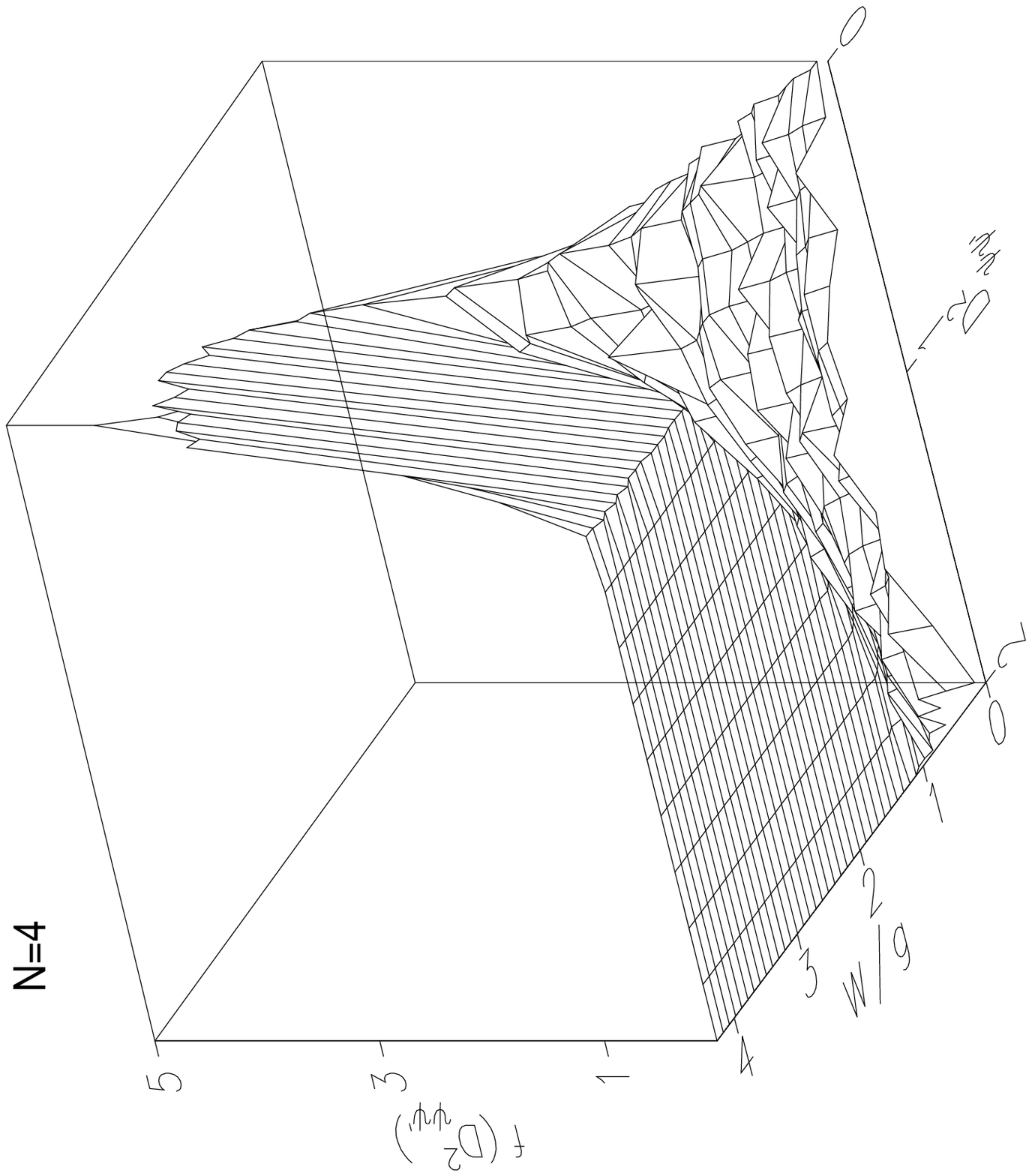}
      }
      \put(8.5,2.5){}
   \end{picture}
\caption[]{Distribution function $f(D^2_{\Psi \Psi'})$ for the 
jump distance of $N$  pseudo spins ($T=0$), neglecting all
entanglements (parameters see Fig.\ \ref{fig2})}
\label{fig5}
\end{figure}

The resulting quantum trajectories are 
characterized as introduced for the exact ones: The respective
distribution function for the jump distance is plotted in Fig.\
\ref{fig5}, where all $M$-terms have been set to zero, a factorization
into individual spins.\\
As we can see, increasing damping rates decrease the difference between
the distributions of the exact and this factorized simulation,
respectively. In the large 
damping regime the latter can be understood 
as a good approximation, which thus allows for
an enormous reduction of the number of relevant state variables. 

\begin{figure}
   \setlength{\unitlength}{1cm}
   \begin{picture}(15,6)
      \put(-2,7){
         \includegraphics{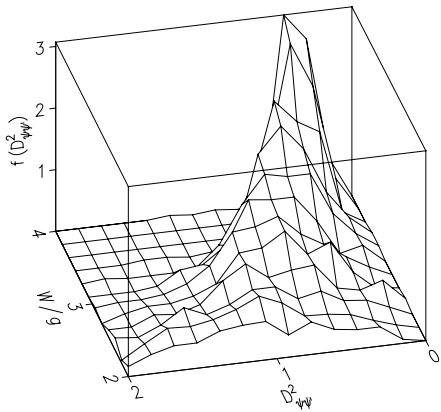}
      }
      \put(8.5,2.5){}
   \end{picture}
\caption[]{Distribution function $f(D^2_{\Psi \Psi'})$ for the 
jump distance of $N=4$  pseudo spins ($T=0$), neglecting all
entanglements beyond second order (parameters see Fig.\ \ref{fig2})} 
\label{fig6}
\end{figure}

\begin{figure}
   \setlength{\unitlength}{1cm}
   \begin{picture}(15,15)
      \put(0,-3.25){
         \includegraphics{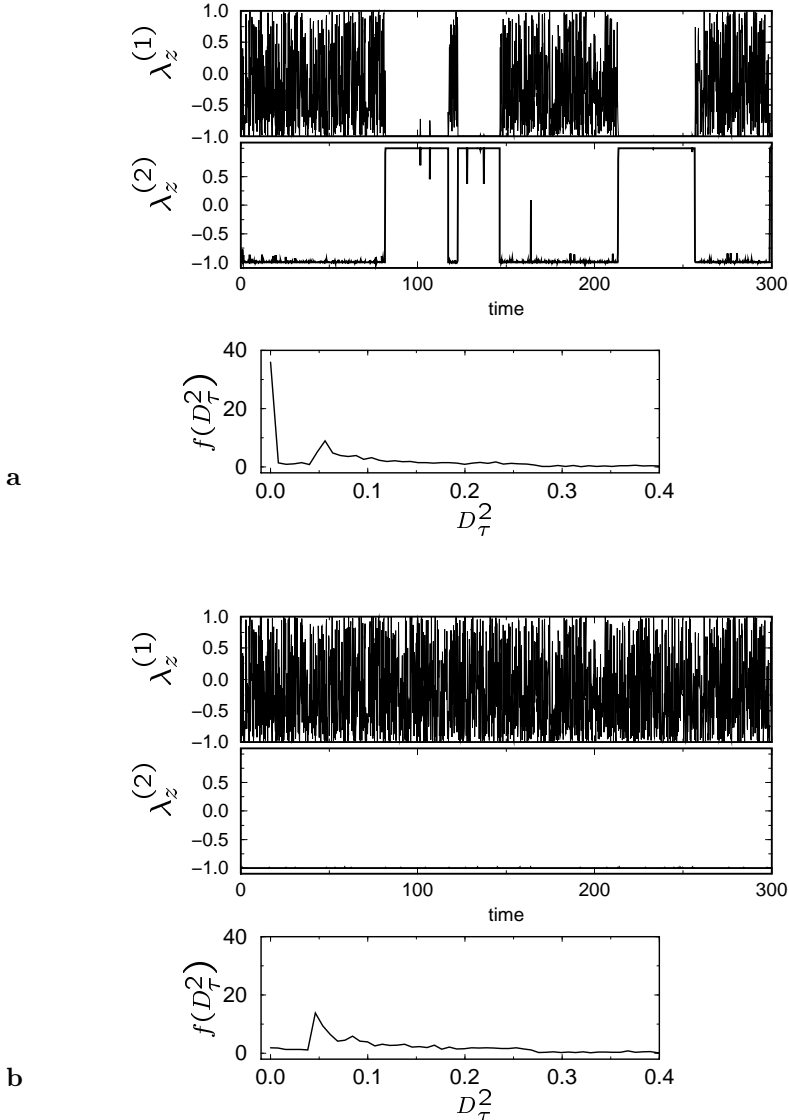}
      }
      \put(0.5,8.5){\bf a}
      \put(0.5,0.5){\bf b}
   \end{picture}
\caption[]{Trajectories of a two spin-system and its distribution
function $f(D^2_{\tau})$ for the exact simulation (a) and  with
neglect of entanglement ((b); parameters: $W_{\downarrow}^{1} =1.0$, 
$g^{1} = 0.7$, $g^{2} = 0.03$, $\delta^{1} = \delta^{2} = -20$,
$C_R = 20$, $\tau = 1.5$ )}
\label{fig7}
\end{figure}

For homogeneous systems adequate factorizations should also be
permutation-symmetric with respect to subsystem-indices. The only
candidates then are truncation schemes. In Fig.\ \ref{fig6} we show an
example for $N=4$ where all entanglement terms beyond second order
($c=2$) are set to zero. Nominally, this approach should constitute an
improvement over the results of Fig.\ \ref{fig5}, for which all
entanglement has been neglected. Qualitatively, the distribution of
Fig.\ \ref{fig6} is, indeed, between the exact and the
non-entanglement-result.\\
However, its significance is difficult to assess, as for $W/g \gg 1$
entanglement becomes negligible, anyway, while for $W/g \to 0$ the
truncation scheme becomes inconsistent. In this region the jump
distance may assume values greater than 2, which has no physical meaning.\\
On the other hand, there are examples in which
the suppression of entanglement can affect quantum trajectories even
in a qualitative way. 
This is demonstrated in Fig.\ \ref{fig7} for an inhomogeneous $N=2$
network. Whereas the exact evolution develops an apparently classical
telegraph  signal between ``light" and ``dark" periods
(``Zenon-effect", \cite{wakelima}), which
appear to be discontinuous only on a large time scale,
these transitions vanish, if entanglement is suppressed!
Fig.\ \ref{fig7} shows a section of the quantum 
trajectories with and without entanglement and the distribution
function of the state distance $D^2_{\tau}$, referring to a
suitable time scale ($\tau \geq 1/W$). The exact simulation has a peak
for $D^2_{\tau} \to 0$, which is missing in the approximation.

\subsection{Co-jump distribution}
\begin{figure}
   \setlength{\unitlength}{1cm}
   \begin{picture}(15,6)
      \put(-2,7){
         \includegraphics{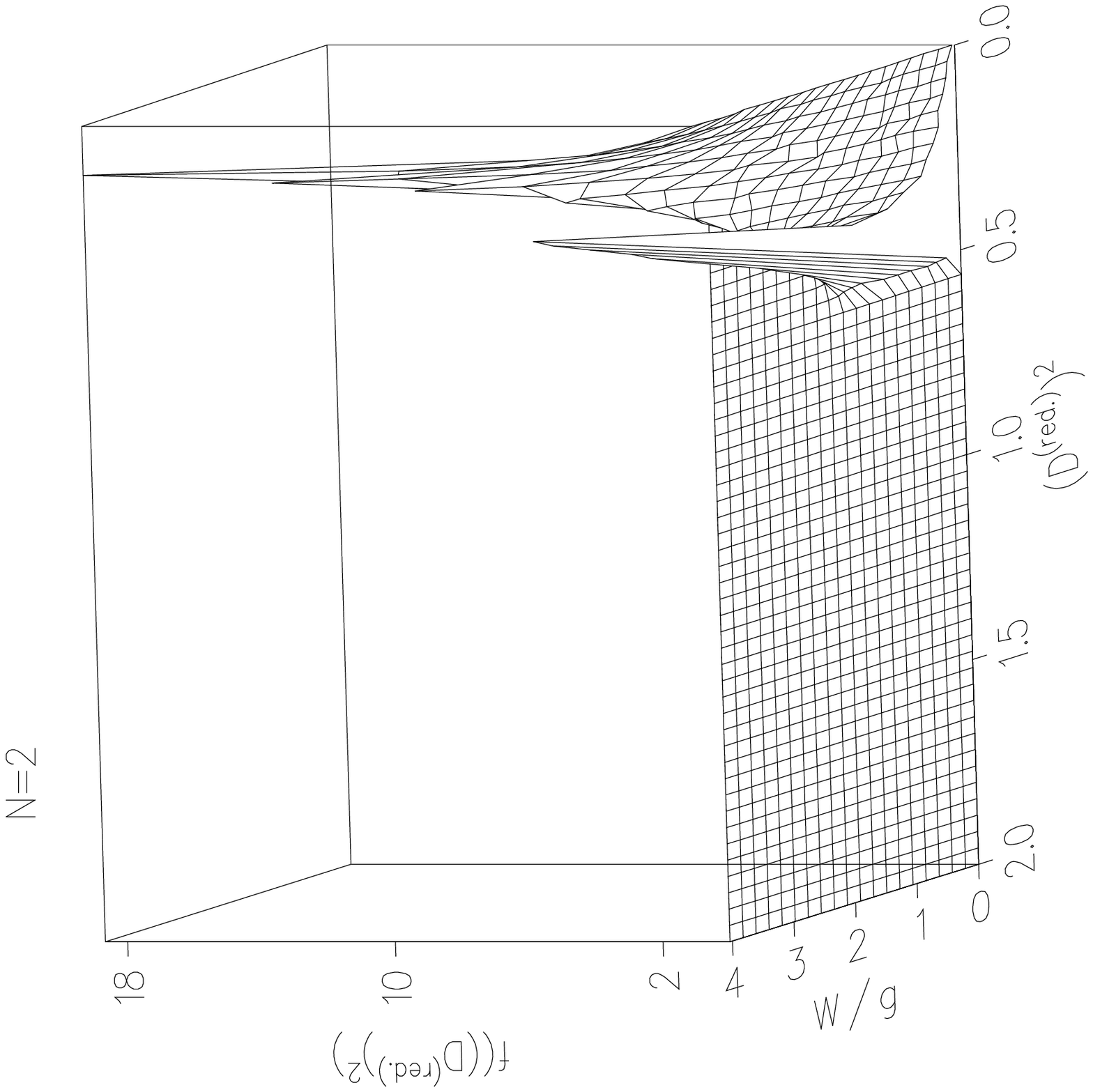}
      }
      \put(6,7){
         \includegraphics{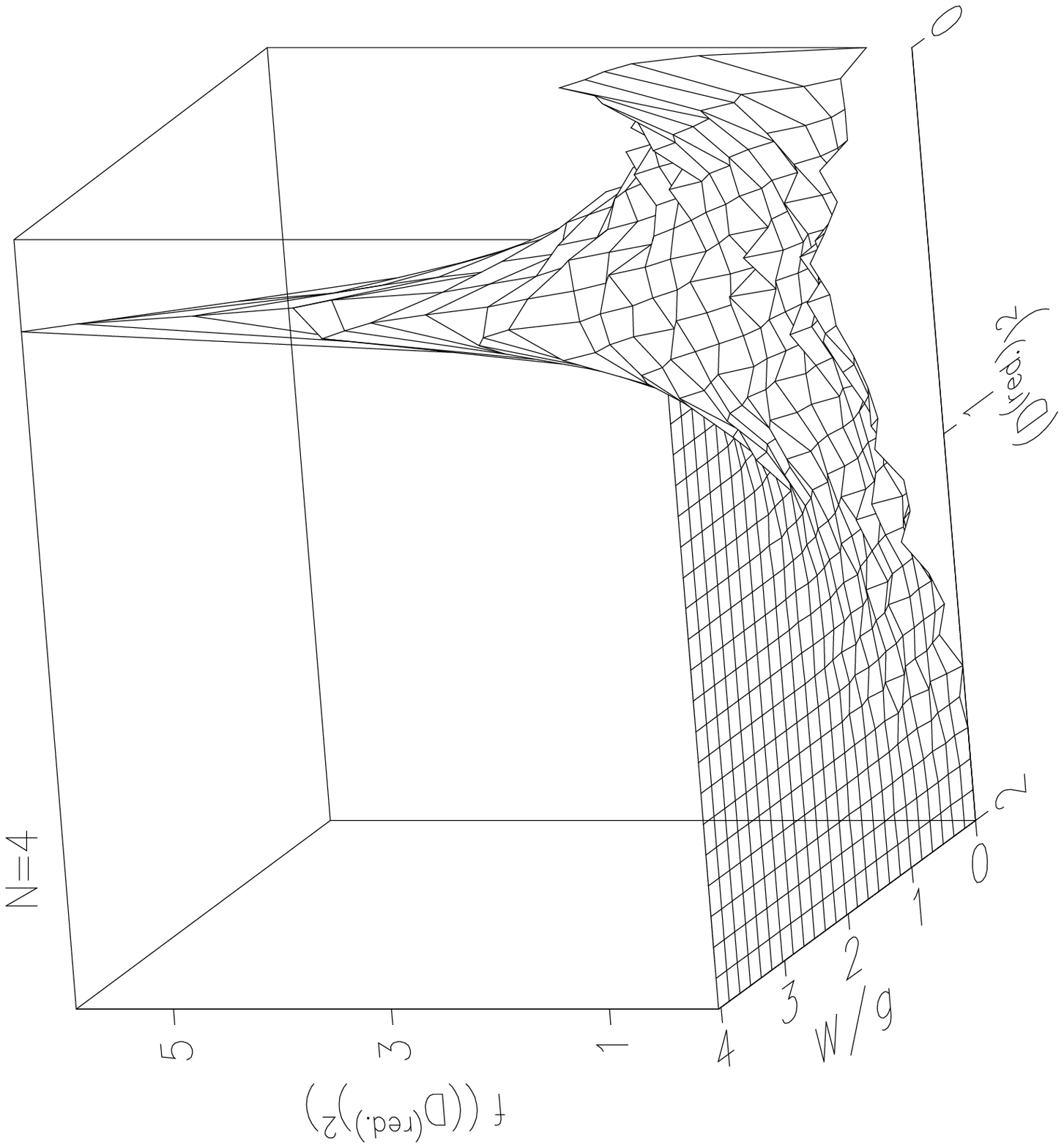}
      }
      \put(8.5,2.5){}
   \end{picture}
\caption[]{Distribution function $f$ for the co-jump distance
$(D^{(red.)})^2$ of $N$  pseudo spins ($T=0$)
(parameters see Fig.\ \ref{fig3})} 
\label{fig8}
\end{figure}

We finally address the non-locality in quantum trajectories:
Using the concept of co-jumps we investigate the
distribution function of the state changes in the respective
non-projected (non 
measured) part of the system. This part changes stochastically from
jump to jump; we therefore denote the respective co-jump-distance
as $D^{(red.)}$. 
Fig.\ \ref{fig8} shows
this entirely non-local effect for $N=2$ and $N=4$. As one can see,
entanglement is still present even in the regime of $W/g
\stackrel{>}{\sim} 4$, 
where the jump- and inter-jump- distribution functions already
indicate a rather ``classical" behavior.  
While the classicality of states implies the classicality of
trajectories, the inverse is not true, in general.
\section{Summary and Conclusions}
We have investigated quantum trajectories of networks with up to $N=4$
subsystems in terms of various statistical state distance
distributions. We 
have found that jump- and inter-jump distributions allow to
describe the classical limit as well as deviations from this limit in
a transparent way. Other measures (like entanglement measures) provide
supplementary information.

The classification of quantum trajectories concerning non-classicality
has to be distinguished from the study of non-classicality of
states. Dynamical properties of open quantum systems
have been investigated in terms of state
changes due to quantum jumps and due to the continuous evolution.
The co-jump distribution may indicate non-classicality of states 
despite the fairly classical trajectories.

For this analysis we found it necessary to abandon the ``ignorance
interpretations" of non-locality, which are often applied in 
quantum-information-scenarios. Non-locality with respect to {\em
individual} pure-state trajectories is, instead, introduced in an operational
way, namely in terms of co-jumps.

In general, large quantum networks do not
necessarily behave ``classical". In the strong damping limit,
``classical trajectories" (``telegraph signals") result (for a bath
temperature $T>0$) which means
that the state-space available to the quantum network is extremely
compressed. This can be exploited for approximation schemes. However,
some caution should be exercised, as truncation schemes
(factorization beyond a certain level $c$), can lead to
inconsistencies for small damping. 

\section*{Acknowledgements}
We thank R.~Wawer, A.~Otte and I.~Kim for valuable
discussions. Financial support by the Deutsche Forschungsgemeinschaft
is gratefully acknowledged.

%
%

%
\end{document}